\documentclass[prb,twocolumn,showpacs,showkeys]{revtex4}
\usepackage{graphicx}
\usepackage{amsmath}
\usepackage{amssymb}
\usepackage{dsfont}

\begin{document}

\title{Relations between Entropies Produced in Nondeterministic Thermodynamic Processes}
\author{S. Turgut}
\affiliation{
Department of Physics, Middle East Technical University,\\
06531, ANKARA, TURKEY}


\begin{abstract}
Landauer's erasure principle is generalized to nondeterministic
processes on systems having an arbitrary number of non-symmetrical
logical states. The condition that the process is applied in the
same way, irrespective of the initial logical state, imposes some
restrictions on the individual heat exchanges associated with each
possible transition. The complete set of such restrictions are
derived by a statistical analysis of the phase-space flow induced
by the process. Landauer's erasure principle can be derived from
and is a special case of these.
\end{abstract}

\pacs{05.20.-y, 89.70.+c}

\keywords{Landauer's erasure principle, Maxwell's demon, nondeterministic processes.}

\maketitle

\section{Introduction}

Landauer's erasure principle (LEP) is a fundamental result in the
thermodynamics of computation. It gives a relation between the
amount of the information stored by a memory device and the
average entropy increase in the environment when a process erases
that information. The principle holds for processes that satisfy
two essential features:
\begin{itemize}
\item[(A)] The process is carried out in the same way independent
of the initial logical state (i.e., the information stored) or the
microstate of the device and

\item[(B)] it restores the device to a known \emph{standard state}
at the end.
\end{itemize}
The first feature is necessary for a complete erasure of
information. Although it is conceivable that the process can read
the information and take different actions depending on it, this
can be done only by recording the information somewhere else. In
that case, the process must also erase the recorded information.
Even when this is the case, the statement (A) remains valid when
the recording instrument is considered as part of the device.

Landauer has shown that under both of these conditions, heats must
be dumped to the environment in such a way that the average
increase of the entropy is at least $k_B\ln2$ per bit of the
information erased,\cite{landauer} where $k_B$ is Boltzmann's
constant. The original derivation of the principle by Landauer
also assumes that the logical states of the device are symmetric,
i.e., they have equal values for the thermodynamical quantities.
However, such a restriction is not necessary when the total
entropy change of both the device and the environment is
considered. For example, for devices having non-symmetrical
logical states, it is seen that the main conclusion of LEP
continues to be valid when a complete write-erase cycle is
considered.\cite{barkeshli}

The principle is of prime importance in Bennett's resolution of
Maxwell's demon paradox.\cite{bennett} In its quest to reduce
entropy, the demon periodically makes measurements, records the
results into its memory and takes some actions depending on these
results. As the memory has a finite capacity, it must be cleared
at some stage and reset to some standard state in order to repeat
the same cycle of events. It is this erasure process and the
associated entropy increase that exactly offsets the reduction of
entropy the demon has achieved and thus saves the second law.
Independently from Landauer and Bennett, same conclusions have
been obtained by Penrose in his textbook on statistical
mechanics.\cite{penrose} A review of Landauer's principle and
Maxwell's demon, as well as reprints of some key articles can be
found in Ref. \onlinecite{md2}.

Since Landauer's original derivation of LEP relies on the second
law where the definition of entropy is extended to an ensemble of
memory devices, several objections have been raised on its
validity. This prompted the appearance of alternative proofs of
the principle that does not rely on the second law, by using the
Fokker-Planck equation\cite{proof1}, by using the microstate
distribution functions,\cite{proof2} and by analyzing the
phase-space flow the erasing process induces.\cite{jacobs}

In all of these, the feature (A) of the process is of central
importance. Since the device and the environment are subjected to
the same treatment independent of the initial logical state,
feature (A) is equivalent to the statement that the
time-dependence of the microstates during the process is governed
by a single, logical-state-independent Hamiltonian. As a result,
the microstates before and after the process are related by a
\textit{single} map. That map is a canonical transformation for
classical systems, which preserves the phase-space volumes by
Liouville's theorem, and is an isometry for quantum systems, which
preserves the dimensions of the subspaces it is acting to. LEP
follows from the constancy of the phase-space measures: as the
process necessarily reduces the phase space of the device by
requirement (B), it must expand that of the environment which
leads to the Landauer bound.

The purpose of this article is to investigate the full
implications of the feature (A) in the most general setting,
without any additional assumptions. It contains the derivation of
a complete set of relations that captures \emph{all} restrictions
that can be placed on the heats dumped to the environment in such
processes. When deriving these relations, no further assumptions
are made on the process and the logical states of the system. The
logical states might be non-symmetrical and, in addition, the
initial and final sets of these states might be different both in
number and in character. Also, the process applied can be
nondeterministic, i.e., for the same initial logical state, it
might lead to a set of different final states with known
transition probabilities. The precise statement that will be proved is the following,\\
\textbf{Theorem 1.} \textit{Suppose that an arbitrary process
having feature (A) is applied on a system that has $n$ initial
logical states and brings it to a final configuration with $m$
logical states. If $P(\beta\vert\alpha)$ denotes the conditional
probability that the initial logical state $\alpha$ ends up in the
final logical state $\beta$, and $k_Bs_{\beta\alpha}$ denotes the
total entropy increase of the system and the environment when an
$\alpha$ to $\beta$ transition occurs, then the following
inequalities are satisfied
\begin{equation}
  \sum_{\alpha=1}^n P(\beta\vert\alpha) e^{-s_{\beta\alpha}} \leq  1\quad (\beta=1,2,\ldots,m)\quad,
\label{eq:CORE}
\end{equation}
i.e., $m$ inequalities, one for each possible final state
$\beta$.}

No additional assumptions other than feature (A) will be made in
proving this theorem. If the process obtains some information
about the system and stores it into some recording instrument or,
in general, an interaction occurs with an instrument, then the
instrument should be considered as part of the system. Apart from
this, if the initial logical state of the system is recorded
elsewhere before the process begins, the conclusions of the
theorem will not change as long as the process does not use this
information.

These relations imply LEP as a special case (when feature (B) is
also imposed), but they go far beyond LEP in implications because
they cannot be derived starting from this principle. It will also
be shown that these relations are complete, i.e., one cannot find
any further restrictions between the transition probabilities and
individual entropy increases. Consequently, any process with
parameters satisfying (\ref{eq:CORE}) can be constructed in
principle. Therefore, these inequalities provide a complete
description of the thermodynamics of classical information
processing by taking nondeterministic logical operations into
account. Like LEP, these relations are mainly of theoretical
interest, but they may find application in the investigation of
heat exchange requirements of probabilistic Turing machines. The
heat exchanges for nondeterministic processes have previously been
analyzed by Penrose.\cite{penrose} They are also investigated in
detail by Maroney for the purpose of generalizing
LEP.\cite{maroney-abs,maroney-gen} However, as it is shown in this
article, their entropic restrictions on heat exchanges are weaker
than theorem 1.

The organization of the article is as follows. In section
\ref{sec:REL}: first, the notation used for describing the systems
and the processes is defined. Then, concentrating on the constant
temperature case, the proof of theorem 1 is given for both
classical and quantum systems. Finally, the completeness of these
relations is proved by constructing an arbitrary process on a
simple system. In section \ref{sec:discussion}, the connection
between the relations (\ref{eq:CORE}) and previously known
relations such as LEP is discussed. Section \ref{sec:conclusion}
contains a brief conclusion. Finally, Appendix \ref{app:vary}
contains the proof of the theorem for the case of various baths at
different temperatures.

\section{Relations between Heats Dumped}
\label{sec:REL}

\subsection{Definitions and Notation}
\label{subsec:notation}

Consider a system $S$ which can have two possibly identical
configurations which will be called \emph{initial} and
\emph{final} configurations. In the initial configuration, the
system has $n$ distinct logical states described by a Hamiltonian
$H_{iS}$ and is in equilibrium with a heat bath $B_i$ at
temperature $T_i$. In the final configuration, the system has $m$
distinct logical states described by Hamiltonian $H_{fS}$ and is
in equilibrium with a bath $B_f$ at temperature $T_f$. The two
baths might be identical. The system can be microscopic, but all
of the baths must be macroscopic.

A process brings the system from the initial configuration to the
final configuration, while in the meantime it brings the system
into contact with various heat baths $B_j$ at different
temperatures $T_j$. Any process satisfying the feature (A) can be described
as a \emph{time-dependent Hamiltonian}
\begin{equation}
\mathcal{H}(t)=H_S(\mathbf{s};t)
+H_{int}(\mathbf{s};\mathbf{b}_i,\mathbf{b}_2,\ldots,\mathbf{b}_f;t)
+\sum_j H_{B_j}(\mathbf{b}_j)
\end{equation}
where $\mathbf{s}$ and $\mathbf{b}_j$ denote the phase-space
coordinates of the system $S$ and the bath $B_j$ respectively,
$H_{B_j}$ is the Hamiltonian of bath $B_j$ and $H_{int}$ describes
the detailed coupling of the system to the baths. If the process
starts at time $t_i$ and ends at $t_f$, the total Hamiltonian
reduces to the corresponding expressions for each configuration.
In other words, for $t\leq t_i$,
\begin{equation}
\mathcal{H}(t)=H_{iS}(\mathbf{s})+H_{i,int}(\mathbf{s};\mathbf{b}_i)+\sum_j H_{B_j}(\mathbf{b}_j)\quad,
\end{equation}
and for $t\geq t_f$
\begin{equation}
\mathcal{H}(t)=H_{fS}(\mathbf{s})+H_{f,int}(\mathbf{s};\mathbf{b}_f)+\sum_j H_{B_j}(\mathbf{b}_j)\quad.
\end{equation}
In both of these configurations, it will be assumed that the
coupling terms $H_{c,int}(\mathbf{s},\mathbf{b}_c)$ ($c=i,f$) are
negligibly small. If $H_{f,int}$ is identically zero, then
sufficiently long times should pass during the process for an
effective equilibration.

The phase space of the system is divided into disjoint regions
corresponding to each logical state in such a way that any
microstate $\mathbf{s}$ belongs to one and only one logical state.
This division can be described with indicator functions
$\theta_{c\alpha}$ which are defined as
$\theta_{c\alpha}(\mathbf{s})=1$ when $\mathbf{s}$ belongs to
logical state $\alpha$ in configuration $c$ and
$\theta_{c\alpha}(\mathbf{s})=0$ otherwise. As a result,
$\sum_\alpha\theta_{c\alpha}(\mathbf{s})=1$ must be satisfied for
all $\mathbf{s}$.

Miscellaneous thermodynamical quantities for each logical state
must be defined in the canonical ensemble by using only those
microstates that belong to the given logical state. Thus, using
the indicator functions, the partition function and the internal
energy are
\begin{eqnarray}
  Z_{c\alpha} &=& \int \theta_{c\alpha}(\mathbf{s}) e^{-H_{cS}(\mathbf{s})/k_BT_c}  d\mathbf{s}\quad, \\
  U_{c\alpha} &=& \frac{1}{Z_{c\alpha}}\int H_{cS}(\mathbf{s})\theta_{c\alpha}(\mathbf{s}) e^{-H_{cS}(\mathbf{s})/k_BT_c}  d\mathbf{s}\quad,
\end{eqnarray}
respectively and the free energy, $F_{c\alpha}=-k_BT_c\ln
Z_{c\alpha}$, and entropy,
$S_{c\alpha}=(U_{c\alpha}-F_{c\alpha})/T_c$, are defined
accordingly.

The logical states must be sufficiently stable in such a way that
they can be used for information storage purposes, i.e., once the
system is in one of the logical states, either it does not make a
transition into another logical state, or the transition time
scales are long compared to the process and equilibration time
scales. In the former case, impenetrable barriers separate the
microstates of different logical states and the above quantities
correspond to the exact canonical thermodynamical functions. In
the latter case, high thermal or diffusion barriers with long
transition times separate the logical states and there is some
arbitrariness in the choice of the indicator functions. Once this
choice is made however, the thermodynamical functions must be
defined as above.

In the proof, the canonical map between the microstates of the
composite system of $S$ and the baths for two given times $t_1$
and $t_2$ will be investigated ($t_1\leq t_i<t_f\leq t_2$). It
will be supposed that the system is prepared such that, at time
$t_1$ it is in logical state $\alpha$ of the initial configuration
and in equilibrium with the bath $B_i$. The initial distribution
of the microstates of the composite system depends on the
preparation, but it must be consistent with the equilibrium
assumption. Consider the microstate of the composite system at
time $t_2$ after the process is applied. Let $P(\beta\vert\alpha)$
be the probability that the final microstate belongs to logical
state $\beta$ of final configuration. Let
$W(\beta\leftarrow\alpha)$ denote the \emph{average} work done
given that an $\alpha$ to $\beta$ transition occurs. This is the
conditional average of the total energy change of the system and
the baths over all microstates that take part in $\alpha$ to
$\beta$ transition. Similarly let $Q_j(\beta\leftarrow\alpha)$
denote the \emph{average} amount of heat transferred to the bath
$B_j$ given that $\alpha$ to $\beta$ transition occurs. As these
quantities must be computed in equilibrium, $t_2-t_f$ must be
sufficiently large for getting thermal equilibrium at time $t_2$.
Moreover, $t_i-t_1$ must be sufficiently large for thermal
equilibration. Provided that both of these conditions are
satisfied, the conditional averages of the initial and final
energies of the system over microstates that take part in $\alpha$
to $\beta$ transition are $U_{i\alpha}$ and $U_{f\beta}$.
Therefore, the quantities above are related by the first law,
\begin{equation}
U_{f\beta}-U_{i\alpha} =
W(\beta\leftarrow\alpha)-\sum_jQ_j(\beta\leftarrow\alpha)\quad.
\end{equation}
The total entropy change per $k_B$ in $\alpha$ to $\beta$
transition will be denoted by $s_{\beta\alpha}$ and is given by
\begin{equation}
  k_B s_{\beta\alpha} = S_{f\beta}-S_{i\alpha}
  +\sum_j\frac{Q_j(\beta\leftarrow\alpha)}{T_j}\quad.
\label{eq:def_s}
\end{equation}
In order to interpret these quantities as the total entropy
change, the baths must be sufficiently large so that the heats
dumped $Q_j(\beta\leftarrow\alpha)$ cannot change their
temperature.

Before going further, it is worth to consider an important special
case, the constant-temperature case, where the system is in
contact with a single bath at temperature $T$ (where $B_i=B_f$ and
$T_i=T_f=T$). In that case, the quantities $s_{\beta\alpha}$ can
be given a simple meaning in terms of the heat emitted to the bath
or the work done on the system. These quantities can be expressed
in terms of $s_{\beta\alpha}$ as
\begin{eqnarray}
  W(\beta\leftarrow\alpha) &=& F_{f\beta}-F_{i\alpha} + k_B T  s_{\beta\alpha} \quad,  \label{eq:Workdone_excessheat}  \\
  Q(\beta\leftarrow\alpha) &=& -T(S_{f\beta}-S_{i\alpha}) + k_B T  s_{\beta\alpha}  \quad.
\label{eq:heatemitted}
\end{eqnarray}
Note that, when some other thermodynamically reversible deterministic process brings
the system from $\alpha$ to $\beta$, $-T(S_{f\beta}-S_{i\alpha})$
is the exact amount of heat that must be dumped to the bath. For
the current nondeterministic process however, the heat dumped
exceeds that reversible contribution by $k_B T s_{\beta\alpha}$.
For this particular reason, for the constant temperature case, it
is tempting to call $s_{\beta\alpha}$ as the dimensionless
\emph{excess heat} (in units of $k_BT$) associated with this
particular transition. For the same reason, it can also be called
as the dimensionless excess work. Despite what the name may imply,
the excess heats can be negative for nondeterministic
transitions.

The excess heats are convenient quantities to be concentrated on,
because in a cyclic change where a set of processes bring the
system back to the initial configuration and state, all reversible
contributions in (\ref{eq:heatemitted}) add up to zero. The sum of
excess heats then gives the total heat dumped to the bath. This is
the case in the context of Maxwell's demon, for example. In this
way, the inconvenience brought by the asymmetry of the states is
eliminated.

\subsection{Proof of Theorem 1 for the Classical Case}
\label{subsec:classical_proof}

This subsection contains the proof of theorem 1 for a classical
system. To make the derivation as clear as possible, it is assumed
that the system interacts with a single bath $B$ at temperature
$T$. The proof of the general case, which involves various baths
at different temperatures, is not different from the proof given
below, but the notation is more involved. For this reason, the
sketch of the general proof is given in the Appendix.

The proof relies on the following approximations which are
justified by the largeness of the bath. (1) At the initial
preparation stage, the total energy $E$ of the composite system
$S+B$ has an arbitrary distribution in a certain range of
energies, say between $E_{min}$ and $E_{max}$. For any $E$ in this
interval, the associated microcanonical temperature of composite
system is approximately $T$. (2) The width of the range,
$E_{max}-E_{min}$, is much larger than the typical energies for
the system. (3) Given that the initial energy of $S+B$ at time
$t_1$ is $E$, the process parameters $P(\beta\vert\alpha)$,
$W(\beta\leftarrow\alpha)$ and $Q(\beta\leftarrow\alpha)$ of the
process is approximately independent of $E$ for $E_{min}\leq E\leq
E_{max}$. The errors made from these approximations will get
smaller as the size of the bath increases. The last assumption
enables us to investigate the relationship between these
quantities within the microcanonical ensemble formalism. In other
words, it is supposed that the initial states of the composite
system has total energy $E$ within the aforementioned range, and
the microstates are distributed with equal \textit{a priori}
probabilities. The basic method is to express the transition
probabilities in terms of certain disjoint phase-space volumes. A
particular addition of these volumes leads directly to the
inequalities of theorem 1.

The process induces a canonical transformation $\phi$ on the total
phase space of $S+B$ from $t_1$ to $t_2$. Let $\phi$ map the
initial points $(\mathbf{s},\mathbf{b})$ to the final points
$(\mathbf{s}^\prime,\mathbf{b}^\prime)=\phi(\mathbf{s},\mathbf{b})$.
Since $\phi$ is canonical, it preserves volumes by Liouville's
theorem, i.e., the phase-space volume elements are equal,
$d\mathbf{s}d\mathbf{b}=d\mathbf{s}^\prime d\mathbf{b}^\prime$.

Let $n_B(E)dE$ represent the phase-space volume of the bath
consisting of points having energy between $E$ and $E+dE$. The
corresponding density is
\begin{equation}
  n_B(E)=\int \delta(E-H_B(\mathbf{b}))d\mathbf{b}\quad.
\end{equation}
As a result of the assumptions made above, the relation
\begin{equation}
  n_B(E-\Delta E)=n_B(E)e^{-\Delta E/k_BT}
\end{equation}
holds with good accuracy for typical energies $\Delta E$ of the
system.

Let $N_{c\alpha}(E)dE$ represent the volume of the total phase
space of $S+B$ of those microstates having energy between $E$ and
$E+dE$ and belonging to state-$\alpha$ of system $S$ in
configuration $c$. The corresponding density is
\begin{eqnarray}
  N_{c\alpha}(E)&=&\int d\mathbf{s}d\mathbf{b}\delta(E-H_{cS}(\mathbf{s})-H_B(\mathbf{b}))\theta_{c\alpha}(\mathbf{s}) \nonumber\\
    &=& \int d\mathbf{s} n_B(E-H_{cS}(\mathbf{s}))\theta_{c\alpha}(\mathbf{s})\nonumber \\
    &=& n_B(E)\int d\mathbf{s} \exp(-H_{cS}(\mathbf{s})/k_BT)\theta_{c\alpha}(\mathbf{s})\nonumber \\
    &=&n_B(E) Z_{c\alpha}\quad.
\label{eq:classical_sys_dos}
\end{eqnarray}
In here, the interaction terms representing bath-system coupling
are dropped assuming that they are negligible.

Finally, let $M_{\beta\alpha}(E^\prime,E)dE^\prime dE$ be the
volume of phase-space points of $S+B$ that start at state-$\alpha$
with energy between $E$ and $E+dE$ and end up at state-$\beta$
with energy between $E^\prime$ and $E^\prime+dE^\prime$. The
corresponding density can be expressed as
\begin{eqnarray}
  M_{\beta\alpha}(E^\prime,E) = \int d\mathbf{s}d\mathbf{b}
        \delta(E-H_{iS}(\mathbf{s})-H_B(\mathbf{b})) \theta_{i\alpha}(\mathbf{s}) \nonumber \\
     \times  \delta(E^\prime-H_{fS}(\mathbf{s}^\prime)-H_B(\mathbf{b}^\prime))
     \theta_{f\beta}(\mathbf{s}^\prime)~.\quad
\label{eq:def_M}
\end{eqnarray}
Note that $M_{\beta\alpha}(E^\prime,E)$ represents the density of
points where an $\alpha$ to $\beta$ transition has occurred and
the total energy has increased by $E^\prime-E$. In that case,
$E^\prime-E$ is the total work done.

Now, suppose that the total energy of $S+B$ was $E$ and the system
was in state $\alpha$ before the process is applied. Then, by
equal \emph{a priori} probabilities assumption, $S+B$ is in one of
the microstates consistent with these restrictions with equal
probability. In that case, the probability that a transition to
state $\beta$ occurs with final energy between $E^\prime$ and
$E^\prime+dE^\prime$ can be expressed as
\begin{equation}
  \frac{M_{\beta\alpha}(E^\prime,E)dE^\prime}{N_{i\alpha}(E)}\quad.
\end{equation}
Let $P(\beta\vert\alpha;E)$ be the probability of making an
$\alpha$ to $\beta$ transition, irrespective of the work done, and
let $\mathcal{P}_{\beta\alpha}(w;E)$ be the probability
distribution function for the work done $w$ in the $\alpha$ to
$\beta$ transition. These probabilities can be expressed as
\begin{eqnarray}
  P(\beta\vert\alpha;E)  &=& \int \frac{M_{\beta\alpha}(E^\prime,E)dE^\prime}{N_{i\alpha}(E)}\quad, \\
  \mathcal{P}_{\beta\alpha}(w;E)    &=& \frac{M_{\beta\alpha}(E+w,E)}{P(\beta\vert\alpha;E) N_{i\alpha}(E)} \quad.
\end{eqnarray}
As discussed above, both of these probabilities have only a weak
dependence on $E$, which can be considered to be a dependence on
the temperature $T$. For this reason, we will write
$P(\beta\vert\alpha;E)=P(\beta\vert\alpha)$ and
$\mathcal{P}_{\beta\alpha}(w;E)=\mathcal{P}_{\beta\alpha}(w)$; the
dependence on the temperature is assumed, but not shown
explicitly. Combining these with (\ref{eq:classical_sys_dos}) we
get
\begin{equation}
  M_{\beta\alpha}(E^\prime,E)=n_B(E)Z_{i\alpha}P(\beta\vert\alpha)\mathcal{P}_{\beta\alpha}(E^\prime-E)\quad.
\label{eq:Expr4M}
\end{equation}
This equation relates the phase-space volume densities to the
process dependent quantities: the transition probabilities and the
distribution function for the work done in individual transitions.

To obtain the inequalities of the theorem, we sum (\ref{eq:def_M})
over the initial logical states and integrate over $E$ which gives
\begin{eqnarray}
  & \sum_\alpha \int dE M_{\beta\alpha}(E^\prime,E) = \int^\prime d\mathbf{s}^\prime d\mathbf{b}^\prime \qquad\qquad
        \nonumber \\
        &\qquad\qquad
         \times\delta(E^\prime-H_{fS}(\mathbf{s}^\prime)-H_B(\mathbf{b}^\prime)) \theta_{f\beta}(\mathbf{s}^\prime)\quad,
\label{eq:Mintegral}
\end{eqnarray}
where the equality of the volume elements,
$d\mathbf{s}d\mathbf{b}=d\mathbf{s}^\prime d\mathbf{b}^\prime$, is
used and the prime on the integral sign indicates that the
integration is over all possible final microstates. This integral
is not over the whole of the phase space when the process map
$\phi$ is not onto. This may happen when infinite, impenetrable
barriers evacuate some part of the phase space. For example, this
is the case for the erasure process considered by Landauer where
all final logical states other than the standard state are
inaccessible. For this reason, the right hand side of
(\ref{eq:Mintegral}) is smaller than the integral over the whole
of phase space,
\begin{equation}
  \sum_\alpha \int dE M_{\beta\alpha}(E^\prime,E) \le
        N_{f\beta}(E^\prime)=n_B(E^\prime)Z_{f\beta}\quad.
\label{eq:first_ineq}
\end{equation}
This is the first place where an inequality is introduced. The
equality holds if and only if $\phi$ is onto.

Using (\ref{eq:Expr4M}) and changing the integration variable to
$w=E^\prime-E$ gives
\begin{eqnarray}
  \sum_\alpha P(\beta\vert\alpha) \int dw \mathcal{P}_{\beta\alpha}(w) e^{-(w+F_{i\alpha}-F_{f\beta})/k_BT} &=&\qquad\nonumber\\
  \sum_\alpha P(\beta\vert\alpha) \langle e^{-w/k_BT}\rangle_{\beta\alpha} e^{-(F_{i\alpha}-F_{f\beta})/k_BT} &\leq&1~,
\end{eqnarray}
where $\langle\cdots\rangle_{\beta\alpha}$ represents the
conditional average over microstates that take part in $\alpha$ to
$\beta$ transition. Using the strict convexity of the exponential
function we get $\exp(-\langle
w\rangle_{\beta\alpha}/k_BT)\leq\langle\exp(-w/k_BT)\rangle_{\beta\alpha}$
where equality holds if and only if work done has no fluctuations.
Finally, using $\langle
w\rangle_{\beta\alpha}=W(\beta\leftarrow\alpha)$ and the relation
(\ref{eq:Workdone_excessheat}), we get the desired inequality in
Eq.~(\ref{eq:CORE}).$\Box$

Inequalities are introduced at two points; therefore if a process
has equalities for all of these $m$ relations, the process map
$\phi$ on the total phase space should be onto and works done
should have absolutely no fluctuations. We
will define $I_\beta$, the \emph{inefficiency} of the process for
the final state $\beta$, as
\begin{equation}
  e^{-I_\beta} = \sum_\alpha P(\beta\vert\alpha) e^{-s_{\beta\alpha}}\quad.
\end{equation}
The relations (\ref{eq:CORE}) then imply that all inefficiencies
are non-negative. Inefficiencies are essentially a combined
measure of the irreversibility of the process and the fluctuations
in the energy exchanges. As it will be seen below, by process
engineering, some excess heats can be reduced and all
inefficiencies can be made to vanish. In the case of an efficient
process, where all $I_\beta=0$, it is not possible to decrease any
of the excess heats without increasing some other excess heat
corresponding to a transition with the same final state. By the
discussion above, efficient processes have no fluctuation for the
excess heats and the process map is onto.

It might be interesting to view the derivation above from the
perspective of the time-reversed process. The time reversal of a
process is always well defined in the Hamiltonian description; it
is simply given by the Hamiltonian
$\tilde{\mathcal{H}}(t)=\mathcal{H}(-t)$ and brings the system
from the final configuration to the initial one. Basically, one
needs to carry out the same actions in reverse order. Below, the
associated quantities for the reverse process will be indicated by
tildes. For simplicity, consider the case where $\phi$ is onto. As
the reversed process map is $\tilde{\phi}=\phi^{-1}$, the
densities in (\ref{eq:def_M}) are related by
$\tilde{M}_{\alpha\beta}(E,E^\prime)=M_{\beta\alpha}(E^\prime,E)$.
Invoking (\ref{eq:Expr4M}), the following relation between the
probability distributions can be found
\begin{eqnarray}
  \tilde{P}(\alpha\vert\beta)\tilde{\mathcal{P}}_{\alpha \beta}(-w) = P(\beta\vert\alpha)
     \mathcal{P}_{\beta\alpha}(w) e^{-w/k_BT}\frac{Z_{i\alpha}}{Z_{f\beta}}~.
\end{eqnarray}

That relation can be made simpler by expressing it in terms of a
microstate dependent variable for the excess heat,
$s=(w+F_{i\alpha}-F_{f\beta})/k_BT$. Note that $\langle
s\rangle_{\beta\alpha}=s_{\beta\alpha}$ and $s$ has no
fluctuations if and only if $w$ has no fluctuations. This variable
can also be expressed as a function of the microstate coordinates
as
\begin{eqnarray}
 s &=&  \frac{1}{k_BT}\Big(
     \sum_\gamma F_{i\gamma}\theta_{i\gamma}(\mathbf{s})
   - \sum_\gamma F_{f\gamma}\theta_{f\gamma}(\mathbf{s}^\prime)   \nonumber \\
  & & + H_{fS}(\mathbf{s}^\prime)+H_{B}(\mathbf{b}^\prime)
      - H_{iS}(\mathbf{s})-H_{B}(\mathbf{b})   \Big) ~,
\end{eqnarray}
for the forward process. Therefore, for the reversed process with
initial point at $(\mathbf{s}^\prime,\mathbf{b}^\prime)$, the
value of the corresponding variable is $-s$. The relationship
between the probability distributions of $s$ for forward and
reverse processes then becomes
\begin{equation}
  \tilde{P}(\alpha\vert\beta)\tilde{\mathcal{P}}^\prime_{\alpha\beta}(-s)
      = P(\beta\vert\alpha) \mathcal{P}_{\beta\alpha}^\prime(s) e^{-s}\quad,
\label{eq:reverse_forward}
\end{equation}
where $\mathcal{P}^\prime$ denotes the distribution function for
that quantity. The inequalities (\ref{eq:CORE}) are obtained by
using the fact that the total probability for the reversed process
is 1.

\subsection{Sketch of the Proof for Quantum Systems}

For quantum systems, there is a problem involved in the definition
(\ref{eq:def_s}) of $s_{\beta\alpha}$. To provide a consistent
definition of these quantities, it should be assumed that the
system starts from a definite initial logical state $\alpha$ and
when the process is completed a projective measurement of the
final logical state is carried out. Provided that this is done,
the average heats dumped to the baths can be computed from the
expectation value of the respective Hamiltonians of the baths and
therefore $s_{\beta\alpha}$ are well-defined quantities. It is
important that such a final measurement stage takes place to
eliminate the possibility of having final microstates in a
superposition state of various logical states. If this is the
case, the inequalities of theorem 1 are valid.

The proof for this case is not different from the classical proof
given above. There is only a change in the terms used. Instead of
a canonical map, there is now an isometry $V$ that maps the
initial microstates into the final ones, which preserve the
dimensions of the subspaces ($V^\dagger V=\mathds{1}$). As above,
the map $V$ does not need to be onto, i.e., it does not need to be
unitary. In that case, $VV^\dagger$ is a projection operator on
the accessible final states and therefore $VV^\dagger \leq
\mathds{1}$.

The operators $\theta_{c\alpha}$ are now a complete set of
orthogonal projections ($\sum_\alpha \theta_{c\alpha}=\mathds{1}$)
which commute with the respective Hamiltonians $H_{cS}$. All of
the phase-space densities defined above can now be expressed as
\begin{eqnarray}
n_B(E) &=& \mathrm{tr}~ \delta(E-H_B)\quad,
 \label{eq:quantum_dos}\\
N_{c\alpha}(E)&=& \mathrm{tr} \left(\delta(E-H_{cS}-H_B)\theta_{c\alpha}\right)\quad,
 \label{eq:quantum_sys_dos}\\
M_{\beta\alpha}(E^\prime,E) &=& \mathrm{tr}
    \left(V\delta(E-H_{iS}-H_B)\theta_{i\alpha}V^\dagger \right. \nonumber \\
    & & \left.\times \delta(E^\prime-H_{fS}-H_B)\theta_{f\beta}\right)\quad.
\end{eqnarray}
The identity in Eq.~(\ref{eq:Expr4M}) remains the same. The
inequality in Eq.~(\ref{eq:first_ineq}) follows by using the fact
that $VV^\dagger \leq \mathds{1}$ and the same convexity argument
leads to the final proof.

\subsection{Completeness of the Relations in Theorem 1}

The set of inequalities (\ref{eq:CORE}) are also complete. In
other words, one cannot find any more restrictions between the
transition probabilities and entropy increases that cannot be
derived from the given inequalities. This completeness statement
is captured in the following theorem.

\textbf{Theorem 2.} \textit{Let $P(\beta\vert\alpha)$ be some
transition probabilities and $s_{\beta\alpha}$
($\beta=1,\ldots,m$, $\alpha=1,\ldots,n$) be some numbers such
that none of $s_{\beta\alpha}$ are infinite and the quantities
$I_\beta$, which are defined as
\begin{eqnarray}
  \sum_\alpha P(\beta\vert\alpha) e^{-s_{\beta\alpha}} = e^{-I_\beta}\quad,
\end{eqnarray}
are nonnegative. Then, there is a system $S$ having initial and
final configurations with respectively $n$ and $m$ logical states,
and there is a process having feature (A) on that system such that
$P(\beta\vert\alpha)$ are the transition probabilities and
$s_{\beta\alpha}$ are the total entropy increases for the
respective individual transitions.}

\textit{Proof:} The system that will be chosen for this purpose is
essentially Szilard's one molecule gas.\cite{szilard} It is a
classical system which is composed of a single molecule inside a
box with a total volume $V$ and in contact with a heat bath at
temperature $T$. The box is divided by impenetrable walls into $n$
regions having volumes $R_{i\alpha} V$ ($\alpha=1,2,\ldots,n$) for
the initial configuration and into $m$ regions having volumes
$R_{f\alpha} V$ ($\alpha=1,2,\ldots,m$) for the final
configuration. Here $R_{c\alpha}$ are positive numbers with
$\sum_\alpha R_{c\alpha}=1$. The region that the molecule is
located represents the logical state and all of them are separated
by impenetrable barriers.

Although it is not essential for the proof of the theorem, it can
be supposed that different constant potentials are applied to each
region. If the molecule is an ion, this can be achieved by
surrounding each region by a metallic sheet and applying different
constant electrostatic potentials to them. Let $u_{c\alpha}$ be
the potential energy in region-$\alpha$ for configuration $c$.
This potential does not affect the motion of the molecule; its
sole purpose is to adjust the internal energies of states to
different values. The internal energy and entropies of the
molecule can be expressed as
\begin{eqnarray}
   U_{c\alpha} &=& u_{c\alpha} + f(T)\quad,\\
   S_{c\alpha} &=& k_B\ln(R_{c\alpha} V) + g(T)\quad,
\end{eqnarray}
for some functions $f$ and $g$ of temperature. Therefore, by
selecting $u_{c\alpha}$, $R_{c\alpha}$ and the volume $V$
appropriately, it is possible to set the values of internal energy
and entropies to anything that we choose. As it will be seen
below, these values do not have any effect on the end results;
they are completely arbitrary.

Consider the following process applied on this gas. Note that all
of the individual steps of this process can be carried out without
knowing where the molecule is. Therefore, it is a process that
satisfies the condition (A). At each step, only the value of the
work done is computed; the heat exchange with the bath can be
computed from the first law at the final stage.

\begin{figure}
\includegraphics[scale=0.55]{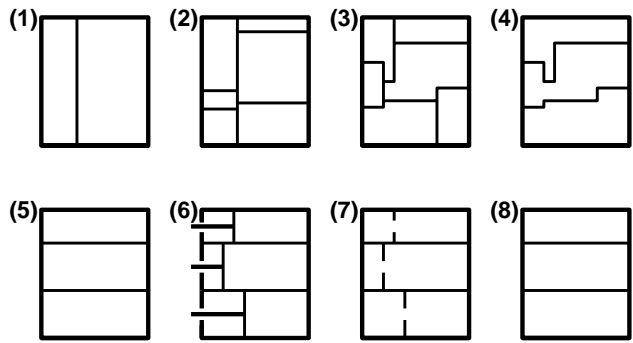}
\caption{The steps of the process on a single molecule system
with $n=2$ initial and $m=3$ final logical states. The system is prepared
in initial configuration and (1) potentials of each region are
lowered to zero, (2) additional walls are inserted, (3) volume of
each region is expanded or compressed, (4) some parts of the walls
are removed, (5) walls are aligned at constant volume, (6) each
region is compressed adiabatically by pistons from left, (7) small
holes opened on the pistons to let the molecule expand freely
towards the left part, and finally (8) pistons are removed and the
potentials of each region are increased.} \label{fig:example}
\end{figure}

Prepare the box in the initial configuration. The molecule
can be in any region.
\begin{itemize}
\item[(1)] First, decrease the potential of each region from
$u_{i\alpha}$ to zero. No heat exchange with the bath occurs in
this step and if the molecule is in region-$\alpha$, then the work
done on it is $W_1=-u_{i\alpha}$.

\item[(2)] Next, divide each region into smaller volumes by
inserting secondary walls, so that region-$\alpha$ (which had
volume $R_{i\alpha}V$) is divided into $m$ smaller regions with
volumes $V_{\beta\alpha}=P(\beta\vert\alpha)R_{i\alpha}V$. There
will be $nm$ such smaller regions at the end, some of which might
have zero volume. This step does not require an expenditure of
work. If molecule was in region-$\alpha$, then the probability
that it will appear in the region-$\beta\alpha$ with volume
$V_{\beta\alpha}$ is $P(\beta\vert\alpha)$.

\item[(3)] Next, slowly change the volume of each of these $nm$
regions such that the volume of region-$\beta\alpha$ is changed
from $V_{\beta\alpha}$ to
\begin{equation}
  V^\prime_{\beta\alpha}=R_{f\beta} P(\beta\vert\alpha) e^{I_\beta-s_{\beta\alpha}}V\quad.
\end{equation}
The work done on the molecule if it is in region $\beta\alpha$ is
\begin{eqnarray}
  W_2 &=& k_BT\ln\frac{V_{\beta\alpha}}{V^\prime_{\beta\alpha}} \\
   &=& k_BT\left(\ln\frac{R_{i\alpha}}{R_{f\beta}}-I_\beta+s_{\beta\alpha}\right)\quad.
\end{eqnarray}
Note that no problem arises if a particular volume
$V_{\beta\alpha}$ is zero, since the probability that the molecule
is there is zero.

\item[(4)] Now, for each $\beta$, connect the regions with indices
$\beta1$, $\beta2$, ..., $\beta n$ into a single connected region
with volume
\begin{equation}
  V^\prime_\beta = \sum_\alpha V^\prime_{\beta\alpha} = R_{f\beta} V\quad,
\end{equation}
by removing some parts of the walls. This step also does not
require an expenditure of work.

\item[(5)] Next, move the walls slowly in such a way that each
region has still the same total volume, but at the end, the walls
become arranged just like the final configuration. This step is
needed for aesthetical reasons only. As long as the volumes of
each region remain the same, there is no work done.

\item[(6)] Now, slowly compress each region-$\beta$ from volume
$V^\prime_\beta$ to
$V^{\prime\prime}_\beta=e^{-I_\beta}V^\prime_\beta$ by a movable
piston. If the molecule is in region-$\beta$, the work done on it
is
\begin{equation}
  W_3 =k_BT\ln\frac{V^\prime_\beta}{V^{\prime\prime}_\beta}=k_BTI_\beta\quad.
\end{equation}

\item[(7)] Now, open a hole on the pistons used in step 6 (or
remove them suddenly), and let the molecule freely expand from
volume $V^{\prime\prime}_\beta$ back to volume $V^\prime_\beta$.
The work done on the molecule is zero again. The steps 6 and 7 are
needed to increase the total entropy by the desired amount. Note that step 7 is possible
only if all $I_\beta$ are non-negative.

\item[(8)] Increase the potential energy of each region from zero
to $u_{f\beta}$. The work done if the molecule is in
region-$\beta$ is $W_4=u_{f\beta}$.
\end{itemize}

At this point, the system is brought to the final configuration.
If the molecule is initially in region-$\alpha$, then it will
appear in region-$\beta$ with probability $P(\beta\vert\alpha)$.
The total work done when this is the case is
\begin{eqnarray}
W(\beta\leftarrow\alpha) &=& \sum_{j=1}^4 W_j \\
        &=&  u_{f\beta}-u_{i\alpha}+k_BT\left(\ln\frac{R_{i\alpha}}{R_{f\beta}}+s_{\beta\alpha}\right)  \\
        &=& F_{f\beta}-F_{i\alpha}+k_BT s_{\beta\alpha}\quad.
\end{eqnarray}
Therefore, the excess heat of that transition is
$s_{\beta\alpha}$.$\Box$

Note that when all expansion and compression steps are done
infinitely slowly there will be no fluctuations in the excess
heats. This implies that the process map $\phi$ is not onto if
some inefficiencies are non-zero. It is obvious that the step 6 is
responsible for this effect as it caused the evacuation of some
part of the total phase-space.

There is a stronger form of this theorem concerned with the
general case where the system interacts with different heat baths
at different temperatures. It states that the inequalities
(\ref{eq:CORE}) capture also all restrictions that are satisfied
by individual heats dumped to the baths. In other words, if each
$Q_j(\beta\leftarrow\alpha)$ are chosen such that the quantity
$s_{\beta\alpha}$ defined in Eq.~(\ref{eq:def_s}) satisfies
(\ref{eq:CORE}), then it is possible to construct a process where
the average heats dumped into each bath in each possible
transition are given by the chosen quantities. The proof of this
statement is not complicated; one only needs to change the step 3
of the process described above in an appropriate way. For this
reason, the construction of the process and the proof of this
stronger statement are left to the reader.

\section{Discussion}
\label{sec:discussion}

Since the inequalities in (\ref{eq:CORE}) express all restrictions
that can be placed on the excess heats, they contain other
powerful relations. In this section, it will be shown that a
number of relations that have been obtained by different
researchers in different contexts can be derived from these
inequalities. The simplicity of these derivations is an indication
of the power of these inequalities. A few other implications of
these relations are also discussed at the end.

\subsection{Penrose's Lower Bounds on Excess Heats}

As each term in (\ref{eq:CORE}) has to be less than 1, the
following lower bound for the excess heats in terms of the
corresponding transition probabilities can be given
\begin{equation}
  s_{\beta\alpha} \ge \ln P(\beta\vert\alpha)\quad.
\label{eq:Penrose_lowbound}
\end{equation}
Hence, some excess heats can be negative if the corresponding
transition probabilities are less than one. This inequality has
been first obtained by Penrose in his treatment of
nondeterministic processes.\cite{penrose} Inequalities in
(\ref{eq:CORE}) enable us to see how the lower bound above can be
accomplished. If, $\alpha$ is the only initial state that leads to
the final state $\beta$, then this lower bound can be achieved. If
possible initial states are more than one, then it is not possible
to achieve the lower bound, but it is possible to approach
arbitrarily close to it, at the expense of increasing
$s_{\beta\lambda}$ for all $\lambda\neq\alpha$.

A bound on the fluctuations of the excess heats can also be
obtained if they are Gaussian. In that case, the average of
$\exp(-s)$ for an $\alpha$ to $\beta$ transition is
\begin{equation}
  \langle e^{-s}\rangle_{\beta\alpha} = \exp\left(-s_{\beta\alpha}+\frac{1}{2}\Delta s^2_{\beta\alpha}\right)\quad,
\end{equation}
where $\Delta s_{\beta\alpha}$ is the standard deviation of the
fluctuations. The upper bound on the standard deviation can then
be obtained from (\ref{eq:reverse_forward}) as
\begin{equation}
  \Delta s_{\beta\alpha} \le \sqrt{2\left(s_{\beta\alpha} -\ln  P(\beta\vert\alpha)\right)}\quad.
\end{equation}
In other words, the fluctuations are necessarily suppressed if the
lower bound in (\ref{eq:Penrose_lowbound}) is approached. However, if the
fluctuations are not Gaussian, it is not possible to find such
bounds on the standard deviation.

\subsection{Landauer's Erasure Principle}

Consider a Landauer erasure process where all initial states end
up in the same final state, e.g., in state $\beta=1$ (i.e., a
restore-to-1 process). In this case,
$P(\beta\vert\alpha)=\delta_{\beta1}$, all inefficiencies except
$I_1$ are infinite and for the final state $1$ we have
\begin{equation}
  e^{-I_1}=\sum_\alpha e^{-s_{1\alpha}} \le 1\quad.
\label{eq:exp_LEP}
\end{equation}
This relation has been first obtained by Szilard\cite{szilard} in
connection with his membrane model. Although Szilard associated
changes in entropy with the measurement process, a correct
interpretation would connect it to the erasure as discussed in
Ref.~\onlinecite{leff}.

In order to obtain a lower bound on average excess heats, it is
convenient to introduce the Legendre transform of the
information-theoretic entropy function,
$\sigma(\mathbf{p})=-\sum_\alpha p_\alpha\ln p_\alpha$, of a
probability distribution $\mathbf{p}=(p_1,p_2,\ldots,p_n)$. Let
$\mathbf{x}=(x_1,x_2,\ldots,x_n)$ be an $n$-tuplet of real
numbers. The Legendre transform of $\sigma$ is
\begin{eqnarray}
  J(\mathbf{x}) &=& \min_\mathbf{p} \left(\sum_\alpha p_\alpha x_\alpha -\sigma(\mathbf{p})\right)\\
   &=& -\ln \left(\sum_\alpha  e^{-x_\alpha}\right)\quad,
\end{eqnarray}
where the minimum is taken over all possible probability
distributions. The definition above implies that the inequality
\begin{equation}
  \sum_\alpha p_\alpha x_\alpha -\sigma(\mathbf{p}) \ge J(\mathbf{x})
\label{eq:J_sigma_ineq}
\end{equation}
is satisfied for all $\mathbf{p}$ and $\mathbf{x}$. Moreover, the
inequality becomes an equality if and only if
$p_\alpha=\exp(J(\mathbf{x})-x_\alpha)$.

Taking $x_\alpha=s_{1\alpha}$ in (\ref{eq:J_sigma_ineq}) gives
$J(\mathbf{x})=I_1\geq0$ and
\begin{equation}
  \sum_\alpha p_\alpha s_{1\alpha}  \geq   \sigma(\mathbf{p})\quad.
\label{eq:averasure}
\end{equation}
If the system is a memory device and information is coded such
that the state $\alpha$ appears with probability $p_\alpha$, then
$\sigma(\mathbf{p})$ is the measure of information in nats stored
by the device. If the same resetting process is applied on the
device (or on an ensemble of such devices) irrespective of the
state, then the left-hand side of (\ref{eq:averasure}) is the
(ensemble) average of the excess heats. Therefore, this inequality
states LEP. Note that the inequality holds for any probability
distribution $\mathbf{p}$. As a result, it can  be considered as
an infinite set of mathematical relations between $s_{1\alpha}$.
Interpreted in this way, (\ref{eq:averasure}) is equivalent to
(\ref{eq:exp_LEP}).

\subsection{Generalized Landauer Principle}

The derivation above can be repeated for an arbitrary process as
well, which leads to the generalized form of Landauer's principle.
Consider a process with transition probabilities
$P(\beta\vert\alpha)$ and excess heats $s_{\beta\alpha}$. If the
system is prepared in the initial configuration with probability
distribution of states being $p_\alpha$, then
\begin{equation}
 \tilde{p}_\beta =\sum_\alpha P(\beta\vert\alpha) p_\alpha \quad,
\end{equation}
is the probability distribution of final states and
\begin{equation}
 \bar{s}=\sum_{\alpha\beta}s_{\beta\alpha}P(\beta\vert\alpha)p_\alpha\quad,
\end{equation}
is the average excess heat. A lower bound on $\bar{s}$ can be
obtained as follows.

First, consider a single final state $\beta$. If we take
$x_\alpha=[s_{\beta\alpha}-\ln P(\beta\vert\alpha)]$ (for
$\alpha=1,2,\ldots,n$), then it can be seen that $J(\mathbf{x})$
is equal to the inefficiency $I_\beta$. Using the inequality
(\ref{eq:J_sigma_ineq}), we can then write
\begin{equation}
  \sum_\alpha\left[s_{\beta\alpha}-\ln P(\beta\vert\alpha)\right]
  q^\prime_\alpha -\sigma(\mathbf{q}^\prime) \ge J(\mathbf{x})=I_\beta \quad,
\end{equation}
which is valid for all probability distributions
$\mathbf{q}^\prime$. Using the particular distribution
$q^\prime_\alpha =P(\beta\vert\alpha)p_\alpha/\tilde{p}_\beta$ in
this inequality gives
\begin{equation}
  \sum_\alpha \left[s_{\beta\alpha}P(\beta\vert\alpha)p_\alpha +P(\beta\vert\alpha) p_\alpha \ln p_\alpha\right]
        -\tilde{p}_\beta \ln \tilde{p}_\beta \ge \tilde{p}_\beta I_\beta \quad.
\end{equation}
Finally, summing over the final states $\beta$ gives
\begin{equation}
  \bar{s} +\sigma(\tilde{\mathbf{p}})-\sigma(\mathbf{p})\ge \sum_\beta \tilde{p}_\beta I_\beta   \quad.
\end{equation}
As each inefficiency has to be non-negative, we get the desired
result,
\begin{equation}
 \bar{s} \ge \sigma(\mathbf{p})-\sigma(\tilde{\mathbf{p}})\quad.
\label{eq:GLEP}
\end{equation}
This is identical in content with the non-decreasing property of
Penrose's statistical entropy.\cite{penrose} It is termed as the
generalized Landauer principle by
Maroney\cite{maroney-abs,maroney-gen} as it relates the average
excess heat emitted to the environment to the change in the
information-theoretic entropy of logical-state distributions.

As the properties of the process (namely $P(\beta\vert\alpha)$ and
$s_{\beta\alpha}$) are independent of the choice of the
distribution $\mathbf{p}$, it is possible to view (\ref{eq:GLEP})
as an infinite set of inequalities, one for each possible
distribution $\mathbf{p}$, placed on the process. Although they
have clear physical interpretations, these inequalities are weaker
than and not equivalent to the main inequalities of this article
given in (\ref{eq:CORE}). The reason is that, all expressions in
(\ref{eq:GLEP}) is essentially a combination of
$\bar{s}_\alpha=\sum_\beta P(\beta\vert\alpha)s_{\beta\alpha}$,
the final-state average of the excess heat when the initial state
is $\alpha$. For this reason, (\ref{eq:GLEP}) places restrictions
only on the averages $\bar{s}_\alpha$ and not on the individual
excess heats associated with every transition.

Moreover, (\ref{eq:CORE}) and (\ref{eq:GLEP}) imply an opposite
dependence between individual excess heats of transitions. As an
example, consider a process where all inefficiencies are 0, and
one of the excess heats, say $s_{\beta\alpha}$, is desired to be
decreased. Relations (\ref{eq:GLEP}) alone implies that, this can
be achieved by increasing another excess heat with the \emph{same
initial state} (e.g., $s_{\mu\alpha}$ with $\mu\neq\beta$) in such
a way that $\bar{s}_\alpha$ remains the same. However, according
to (\ref{eq:CORE}), this is not possible; $s_{\beta\alpha}$ can be
decreased only by increasing another excess heat with the
\emph{same final state} (e.g., $s_{\beta\lambda}$ with
$\lambda\neq\alpha$). For this reason, the relations
(\ref{eq:GLEP}) are not sufficient for this kind of process
engineering problems.

\subsection{Controlled Processes}

A trivial application of the inequalities in (\ref{eq:CORE}) is to
controlled processes. Here, the process to be applied on a system
$S$ is determined based on the logical state of another system
$C$, the controller; but the controller does not change its
logical state during the process. The controlled-NOT gate is a
well-known example. The restrictions on the heat exchanges with
the environment for such processes can be analyzed simply by
imposing these restrictions for each individual process on $S$. To
be precise, suppose that when the state of the controller is $k$,
the process $A^{(k)}$ is applied on $S$ which has transition
probabilities $P^{(k)}(\beta\vert\alpha)$ and excess heats
$s^{(k)}_{\beta\alpha}$. For this case, the controlled process
applied on the combined system $S+C$ has the following transition
probabilities and excess heats
\begin{eqnarray}
  P(\beta j\vert\alpha k) &=& \delta_{jk}  P^{(k)}(\beta\vert\alpha) \quad,\\
  s_{\beta k,\alpha k} &=& s^{(k)}_{\beta\alpha}\quad.
\end{eqnarray}
The inequalities (\ref{eq:CORE}) for the controlled process then are
\begin{equation}
  \sum_\alpha P^{(k)}(\beta\vert\alpha) e^{-s^{(k)}_{\beta\alpha}}\leq 1\quad,
\end{equation}
which must hold for all $\beta$ and $k$. Therefore, the controlled
process satisfies the inequalities (\ref{eq:CORE}) if and only if
every individual process $A^{(k)}$ applied on $S$ satisfies the
same inequalities.

An immediate application of the result above is to the measurement
processes. In this case $C$ is the system whose state will be
measured and $S$ plays the role of the recording instrument.
Before the measurement, $S$ must be prepared in a standard state,
say $\alpha=1$. The measurement is then a controlled process as
above, where the individual process $A^{(k)}$ changes the state of
$S$ from $\alpha=1$ to $\beta=k$ with certainty. As this is the
only necessary requirement, it is possible to construct $A^{(k)}$
as a deterministic and logically reversible process and choose $s^{(k)}_{k,1}=0$.
This controlled process changes the logical states of $S+C$ from
$(1,k)$ into $(k,k)$, i.e., the state of $C$ has been copied into
$S$, and no excess heat is transferred to the environment in doing
this. This is a simple, but general demonstration of the principle
first stated by Bennett\cite{bennett}, i.e., thermodynamically reversible
measurements can be done provided that the recording instrument is
initialized in a standard state.

\subsection{Some further Bounds and Processes with Doubly Stochastic Transition Probabilities}

The inequalities in (\ref{eq:CORE}) give the relation between
excess heats of the same final state, e.g., it gives a restriction
between $s_{\beta1},\ldots,s_{\beta n}$. The following lower bound
on the largest of these quantities can be easily deduced
\begin{equation}
  \max_\alpha s_{\beta\alpha} \geq \ln\left(\sum_\alpha  P(\beta\vert\alpha)\right)\quad,
\end{equation}
where the equality applies if and only if
$s_{\beta1}=\ldots=s_{\beta n}$. From here, it is possible to show
that the largest excess heat of all transitions,
$\max_{\alpha\beta} s_{\beta\alpha}$, is bounded from below by
$\ln(n/m)$. In other words, at least one excess heat should exceed
that bound. This result, together with a corresponding one for
final-state averages of excess heats, is also
contained in the following proposition.\\
\textbf{Proposition.} \textit{The following are equivalent.\\
(a) $s_{\beta\alpha}\leq\ln(n/m)$ for all $\beta$ and $\alpha$,\\
(b) $\bar{s}_\alpha \leq\ln(n/m)$ for all $\alpha$, where
$\bar{s}_\alpha =\sum_\beta P(\beta\vert\alpha)s_{\beta\alpha}$
denotes the final-state average of the excess heat for the initial
state $\alpha$,\\
(c) $s_{\beta\alpha}=\ln(n/m)$ for all $\beta$ and $\alpha$, all
inefficiencies are $I_\beta=0$ and the transition probabilities
satisfy
\begin{equation}
  \sum_\alpha P(\beta\vert\alpha)=\frac{n}{m}\quad.
\label{eq:P_double_stoch}
\end{equation}
}

\textit{Proof:} As the implications
(c)$\Rightarrow$(a)$\Rightarrow$(b) are trivial, we only need to
show (b)$\Rightarrow$(c). Suppose that (b) holds. First, consider
a fixed $\alpha$. Strict convexity of the exponential function
leads to
\begin{equation}
  \frac{m}{n}\leq e^{-\bar{s}_\alpha} \leq \sum_\beta
  P(\beta\vert\alpha)e^{-s_{\beta\alpha}}
\end{equation}
where the rightmost inequality is an equality if and only
$s_{\beta\alpha}=\bar{s}_\alpha$ for all $\beta$. Next, sum these
inequalities over the initial state $\alpha$ and apply
(\ref{eq:CORE}) to get
\begin{equation}
  m\leq \sum_\alpha e^{-\bar{s}_\alpha} \leq \sum_{\alpha\beta}
  P(\beta\vert\alpha)e^{-s_{\beta\alpha}}\leq\sum_\beta 1 =
  m\quad.
\end{equation}
As the leftmost and rightmost sides of this chain are equal, all
of the individual terms are equal to each other. Therefore, all
inequalities that are used to obtain it must have been equalities
as well. As a result, $s_{\beta\alpha}=\bar{s}_\alpha=\ln(n/m)$
for all $\alpha$ and $\beta$, $I_\beta=0$ for all $\beta$ and
Eq.~(\ref{eq:P_double_stoch}) follows from these.$\Box$

At this point, it is worth to concentrate on a particular special
case, the case where the initial and final configurations of the
system are identical (and hence $n=m$). Most applications, for
example memory elements used in computation, fall under the scope
of this case. A minor result that follows from the proposition for
this case is that for any process there should be a transition
with a non-negative excess heat. Similarly, there should be a
state $\alpha$ for which $\bar{s}_\alpha\geq0$.

The processes that satisfy the conditions of the proposition
have some remarkable properties that should be mentioned. These
are processes that have doubly stochastic transition probabilities
\begin{equation}
  \sum_\alpha P(\beta\vert\alpha)=  \sum_\alpha  P(\alpha\vert\beta)=1\quad.
\end{equation}
In this case, it is possible to construct the process in such a
way that all excess heats are zero which necessarily implies that
the process is also efficient for all final states. The
proposition is stating that any process that has non-positive
excess heats for all transitions should be such a process.

The excess heats of these processes have absolutely no
fluctuations. There are no macroscopic fluctuations because all
excess heats are uniform, $s_{\beta\alpha}=0$, and there are no
microscopic fluctuations because the process is efficient. If the
logical states of the system are also symmetric, i.e., they have
equal equilibrium energy and entropy, then absolutely no heat
exchange occurs with the bath and there is no work done
irrespective of the initial and final state. A contact with a bath
is not even necessary to implement the process.

Moreover, successive application of two such processes yields a
process that has the same feature. For this reason, they might
find application in the implementation of probabilistic Turing
machines. If the individual computation steps of such a machine
have doubly stochastic transition probabilities, then it is
possible to run the whole computation without any heat exchange
with the environment and without any work done (assuming that the
states are symmetric).

Furthermore, any probabilistic Turing machine can be adapted to
have doubly stochastic transitions. Consider, for example, a step
of the computation where a set of memory elements $S$ with $n$
logical states undergoes a logical operation with transition
probabilities $P(\beta\vert\alpha)$, which may not be doubly
stochastic. Let $A$ be an ancillary device having $n$ logical
states, which is initially in state $i=1$. Consider an operation
on $S+A$ which has the following transition probabilities
\begin{equation}
  P^\prime(\beta j\vert \alpha i) = \left\{
    \begin{array}{ll}
    \frac{1}{n}P(\beta\vert\alpha)  & \textrm{if}~i=1, \\
    \frac{(n-r_\beta)}{n^2(n-1)} & \textrm{if}~i\neq1,
    \end{array}\right.
\end{equation}
where $r_\beta=\sum_\alpha P(\beta\vert\alpha)$ and the roman
letters denote the logical states of $A$. It can be seen that
$P^\prime$ is doubly stochastic and $\sum_j P^\prime(\beta
j\vert\alpha 1)=P(\beta\vert\alpha)$. In other words, if $A$
starts in state $1$, then same logical operation is obtained on
$S$. In order to make the whole computation doubly stochastic,
different ancillaries with the same initial state have to be used
at each computation step. This discussion essentially shows that
probabilistic Turing machines with doubly stochastic transition
probabilities can have the same computing power as any other
probabilistic machine, at the expense of using a larger memory
space.

However, it should be kept in mind that probabilistic Turing
machines can also be designed to have deterministic computation
steps where randomness is introduced by an additional input tape
containing random symbols. They are more manageable for
thermodynamically reversible computation since they can be adapted
to have logically reversible computation steps as in Ref.
\onlinecite{bennettrev}. In that case, they will also be able to
erase all intermediate results of the computation, except the
input, the desired output and the random tape, without any heat
exchange. In addition, the symbols on the random tape can also be
generated and subsequently erased without any heat exchange (e.g.,
inserting walls into a container containing single molecule for
generation; removing the walls for erasing). As a result, there is
no problem in doing probabilistic computation in a
thermodynamically reversible way.

\section{Conclusions}
\label{sec:conclusion}

A complete set of inequalities has been obtained that places
restrictions on the excess heats and the transition probabilities
of any nondeterministic processes having feature (A). These
inequalities can be interpreted as a generalized form of both
Penrose's bound (\ref{eq:Penrose_lowbound}) for nondeterministic
processes and Landauer's bound (\ref{eq:exp_LEP}) for resetting
operations. Their unique power comes from the completeness
property. As a result, just like these two classical results, any
relation between excess heats and transition probabilities can be
obtained starting from these inequalities.

\appendix
\section{Proof of Theorem 1 for Varying Temperature Case}
\label{app:vary}

Consider the general case where the process brings the system into
contact with various heat baths at different temperatures. The
notation is described in subsection \ref{subsec:notation} and the
proof follows the same lines of the one given in subsection
\ref{subsec:classical_proof}. The quantum case is assumed because
of its slightly better notation. It is also assumed that the baths
$B_i$ and $B_f$ are distinct; the case where they are identical is
not considered separately as the proof is altered only slightly.
Suppose that there are $N$ baths and the initial and final baths
are $B_i=B_1$ and $B_f=B_N$ respectively.

Let $n_{B_j}(E)$ denote the density of states of bath $B_j$ at
energy $E$ which is defined by Eq.~(\ref{eq:quantum_dos}) and
$N_{c\alpha}(E)$ denote the density of states of $S$ and $B_c$ for
logical state $\alpha$ at configuration $c=i,f$. These are defined
as in Eq.~(\ref{eq:quantum_sys_dos}) for the appropriate bath and
can be expressed as $N_{c\alpha}(E)=n_{B_c}(E)Z_{c\alpha}(T_c)$.

Let $E_1,\ldots,E_N$ be a set of energies chosen in the
appropriate temperature range of the corresponding baths and
$E_1^\prime,\ldots,E_N^\prime$ be another set of energies. To
simplify the notation, the shorthand
$\mathcal{E}=(E_1,\ldots,E_N)$ and
$\mathcal{E}^\prime=(E_1^\prime,\ldots,E_N^\prime)$ will be used.
Consider the following set of microstates of the system and the
baths.
\begin{itemize}
\item[i.] Initially, before the process is applied
\begin{itemize}
\item[(a)] $S$ is in logical state $\alpha$,

\item[(b)] $S+B_i$ has energy in interval $(E_1,E_1+dE_1)$,

\item[(c)] bath $B_j$ ($j\neq 1$) has energy in interval
$(E_j,E_j+dE_j)$;
\end{itemize}
\item[ii.] and after the process is completed
\begin{itemize}
\item[(a)] $S$ is in logical state $\beta$,

\item[(b)] $S+B_f$ has energy in interval
$(E_N^\prime,E_N^\prime+dE_N^\prime)$,

\item[(c)] bath $B_j$ ($j\neq N$) has energy in interval
$(E_j^\prime,E_j^\prime+dE_j^\prime)$.
\end{itemize}
\end{itemize}
The corresponding ``density of states'' for such microstates is
given by
\begin{eqnarray}
&M_{\beta\alpha}(\mathcal{E}^\prime; \mathcal{E}) =  \qquad\qquad\qquad\qquad\qquad \nonumber \\
& \mathrm{tr}~ \left(
  V  \left[
  \theta_{i\alpha} \delta(E_1-H_{iS}-H_{B_i})
    \prod_{j\neq i} \delta(E_j-H_{B_j})
    \right]  V^\dagger \right.
    \nonumber \\
&   \left. \left[\theta_{f\beta} \delta(E_N^\prime-H_{fS}-H_{B_f})
    \prod_{j\neq f} \delta(E_j^\prime-H_{B_j})\right]\right)
    \nonumber \\
\label{eq:App_def_M}
\end{eqnarray}
where $V$ is the isometry corresponding to the time development of
the state from $t_1$ to $t_2$. It satisfies $V^\dagger
V=\mathds{1}$ and $VV^\dagger\leq\mathds{1}$. Note that, $E_1$ and
$E_N^\prime$ include also the energy of the system $S$, while all
the other energies represent only the corresponding bath's energy.

The density of states in (\ref{eq:App_def_M}) can be easily
related to the process dependent probabilities. First, note that
by equal \emph{a priori} probabilities, all microstates satisfying
condition (i) above can be represented by the density matrix
\begin{equation}
  \rho=\frac{1}{\mathcal{Z}} \theta_{i\alpha} \delta(E_1-H_{iS}-H_{B_i})
    \prod_{j\neq i} \delta(E_j-H_{B_j})
\end{equation}
where
\begin{equation}
 \mathcal{Z}=\sum_\beta \int d\mathcal{E}^\prime M_{\beta\alpha}(\mathcal{E}^\prime; \mathcal{E})
      =Z_{i\alpha}(T_1)\prod_{j=1}^N n_{B_j}(E_j).
\end{equation}
As a result, $\rho^\prime=V\rho V^\dagger$ is the density matrix
when the process is completed. Projective measurements by
$\{\theta_{f\beta}\}$ will result in the transition probabilities
\begin{equation}
  P(\beta\vert\alpha)=\mathrm{tr}~\rho^\prime\theta_{f\beta}=\frac{1}{\mathcal{Z}}\int
  d\mathcal{E}^\prime M_{\beta\alpha}(\mathcal{E}^\prime;\mathcal{E})\quad.
\end{equation}
Finally, if collapse to $\beta$-state occurs in that measurement,
the final density matrix is
$\rho^{\prime\prime}=\theta_{f\beta}\rho^\prime\theta_{f\beta}/P(\beta\vert\alpha)$
from which the distribution of final energies of the baths and the
system, i.e., the function
$\mathcal{P}_{\beta\alpha}(\mathcal{E}^\prime-\mathcal{E})$ can be
computed. The quantity defined in (\ref{eq:App_def_M}) contains
this information. As such, it can be expressed as
\begin{equation}
M_{\beta\alpha}(\mathcal{E}^\prime; \mathcal{E})
  =P(\beta\vert\alpha)  \mathcal{P}_{\beta\alpha}(\mathcal{E}^\prime-\mathcal{E})
  Z_{i\alpha}(T_1)\prod_{j=1}^N n_{B_j}(E_j)\quad,
\label{eq:App_M}
\end{equation}
where again it is assumed that these probabilities have a weak
dependence on initial energies $\mathcal{E}$ and it is supposed
that this dependence can be taken as an implicit dependence on the
temperatures.

Now, sum and integration of $M$ over initial state and energies
gives
\begin{equation}
 \sum_\alpha \int d\mathcal{E} M_{\beta\alpha}(\mathcal{E}^\prime;\mathcal{E})
    \leq  Z_{f\beta}(T_N)  \prod_{j=1}^N n_{B_j}(E_j^\prime)\quad,
\end{equation}
where the inequality is introduced taking by into account that
$VV^\dagger\leq\mathds{1}$. Using (\ref{eq:App_M}), the last
inequality can be written as
\begin{eqnarray}
 \sum_\alpha P(\beta\vert\alpha) \int d\mathcal{E}
    \mathcal{P}_{\beta\alpha}(\mathcal{E}^\prime-\mathcal{E})
    \frac{Z_{i\alpha}(T_1)}{Z_{f\beta}(T_N)}
    \prod_j \frac{n_{B_j}(E_j)}{n_{B_j}(E_j^\prime)}
      \nonumber \\
 =\sum_\alpha P(\beta\vert\alpha)\left\langle e^{-s}\right\rangle_{\beta\alpha} \leq 1~,\qquad
\end{eqnarray}
where now the variable $s$ is given by
\begin{equation}
  s=\frac{F_{f\beta}(T_N)}{k_BT_N}-\frac{F_{i\alpha}(T_1)}{k_BT_1}+\sum_{j=1}^N \frac{E_j^\prime-E_j}{k_BT_j}\quad.
\end{equation}
Since
\begin{eqnarray}
  \langle E_1^\prime -E_1\rangle_{\beta\alpha} &=&  Q_i(\beta\leftarrow\alpha)-U_{i\alpha}\quad,\\
  \langle E_N^\prime -E_N\rangle_{\beta\alpha} &=&  Q_f(\beta\leftarrow\alpha)+U_{f\beta}\quad,\\
  \langle E_j^\prime -E_j\rangle_{\beta\alpha} &=&  Q_j(\beta\leftarrow\alpha)\quad (j\neq 1, N)\quad,
\end{eqnarray}
it can be seen that $\langle
s\rangle_{\beta\alpha}=s_{\beta\alpha}$. The inequalities in
(\ref{eq:CORE}) then follow by using the convexity of the
exponential function. The detailed relation in
Eq.~(\ref{eq:reverse_forward}) between the probability
distributions for reversed and forward processes continue to hold
in this case as well.

\end{document}